# *scellop:* A Scalable Redesign of Cell Population Plots for Single-Cell Data

Thomas C. Smits, Nikolay Akhmetov, Tiffany S. Liaw, Mark S. Keller, Eric Mörth, Nils Gehlenborg

Department of Biomedical Informatics, Harvard Medical School, Boston, MA 02115, United States

## Abstract

**Summary:** Cell population plots are visualizations showing cell population distributions in biological samples with single-cell data, traditionally shown with stacked bar charts. Here, we address issues with this approach, particularly its limited scalability with increasing number of cell types and samples, and present *scellop*, a novel interactive cell population viewer combining visual encodings optimized for common user tasks in studying populations of cells across samples or conditions.

**Availability and Implementation:** Scellop is available under the MIT licence at https://github.com/hms-dbmi/scellop, and is available on PyPI (https://pypi.org/project/cellpop/) and NPM (https://www.npmjs.com/package/cellpop). A demo is available at https://scellop.netlify.app/.

**Contact:** Nils Gehlenborg, nils@hms.harvard.edu

**Supplementary information:** Supplementary information is included at the end of this document.

-

# Introduction

Cell population plots are visualizations showing cell types, states, or clusters in a stratified manner, for instance among samples or conditions. These are used for comparing cell types within and between samples. They support analyzing heterogeneity across cell populations, examining present cell types and comparing cell counts, often in different populations, such as disease states [1], locations [2], and subgroups of cell types [3]. In publications, these are often accompanied by dimensionality reduction plots [4]. They can also be used for antibody isotypes in different cells [5]. Cell populations are traditionally shown using stacked bar charts, with samples as individual bars, cell types as colored segments of various lengths corresponding to the number or proportion of cells.

Cleveland & McGill (1984) [6] highlight a crucial issue with comparing multiple segments in stacked bar charts in their study of human perception, noting participants are better at comparing position than length. Shifted segments are especially hard to compare. Only the bottom cell type in a stacked bar chart has the same starting point and can therefore be compared by the position of the top of its segment. Talbot et al. (2014) [7] expand their study to highlight the effect of comparing lengths with separated bars, noting that comparison between separated bars is harder than adjacent bars. Nobre et al. (2024) [8] show that users achieve lower accuracy and take more time for various tasks using stacked bar charts compared to other common chart types.

Now, with larger single-cell atlas studies that have the increasing ability to detect more and rarer cell types across samples and conditions, the challenge of visually comparing cell populations by the size of the segments in the stacked bar chart has become more pronounced, as bar segments are smaller and separation between bar segments in different datasets is increased. Additionally, increasing the number of cell types also requires more colors to distinguish categories. Sultana et al. (2025) [1] include 30 cell types in their cell population plot. The average number of cell types in annotated RNAseq datasets from the Human BioMolecular Atlas Program (HuBMAP) is 33 (see Supplementary Materials). However, using seven or more colors to visually encode categories impacts readability [9], and identification accuracy decreases with more colors [10]. Therefore, an alternative encoding is necessary for scalable population plots.

Here, we evaluate the user tasks and needs for cell population plots. We redesign these plots and introduce *scellop*, a flexible viewer for comparison and communication of cell populations.

# Methods

## Design Considerations

To evaluate the issues with the traditional cell population view (stacked bar chart approach) and gain an understanding of desired features, we performed a user study (N=14) using the cell population plot in the HuBMAP Data Portal Tissue Blocks Comparison as an example (https://hubmapconsortium.github.io/tissue-bar-graphs/) (Figure S1). See Supplementary Materials for details.

The main desired interactions of a visualization of cell populations were normalization, grouping by cell type hierarchy, overview-to-detail navigation, the ability to filter and group by metadata, and showing additional context for cell types and samples. Additional interactions raised were related to cell type sorting and exporting the visualization. The main issues raised were related to the color scheme. Users could not understand what the different colors represented with many samples, and were unable to change the color scheme to better represent the sample distribution. Other raised issues were regarding the amount and granularity of cell types, making it challenging to get an overview of the distributions. With larger numbers of cell types and samples, identifying absent and universally present cell types proves to be challenging. Samples with different cell type granularities, e.g., with cell type annotation performed by different algorithms, are hard to compare. Other potential issues arise from small fractions, which can be hard to identify and compare. Additionally, when the order of cell types is determined by their overall abundance, this order may not reflect the relative cell type proportions in each sample, confusing the viewer when examining a single sample.

We distinguish different groups of user tasks: (1) viewing the structure of a single sample (e.g., what is the most common cell type, what is the proportion of a given cell type, how do the proportions of multiple cell types in the same sample compare), (2) comparing the structures of multiple samples (e.g., how do proportions of a given cell type compare in different samples, in how many samples is a particular cell type identified, what percentage of total cells of all samples does a given cell type contribute), and (3) comparing the structures of multiple samples in relation to their metadata (e.g., what is the most common cell type for a given organ, is there a correlation between proportion of a cell type and sample metadata). In all of these tasks, the ability to show cell type hierarchy and group and filter on these is imperative.

To support these user tasks, we developed an interactive tool called scellop. We redesign cell population plots to better support cell type comparisons. We use a central heatmap for general trends, encoding samples and cell types as rows and columns. Several tools exist for heatmap-like views with flexible encodings (Bertifier [11], Clustergrammer [12], Funkyheatmap [13]), building on Bertin's matrix principles [14]. Although these can show overall patterns and help users to compare trends in multiple samples, they do not support inspection of individual sample structures, various normalization and transformation operations (except Clustergrammer), and operations on cell type hierarchies. In scellop, to allow for within-sample and between-sample comparisons, each heatmap row can be expanded into a bar chart. Bar and violin plot panels aligned to the heatmap display cell counts and distributions. Scellop supports all desired interactions identified from our design study, including normalization, grouping and filtering. Together, these features comprise the full set of tasks from Schneiderman's task taxonomy for information visualization: *overview - zoom - filter - details-on-demand - relate - history - extract* [15].

## Implementation

Scellop is available as a Python package on PyPI (https://pypi.org/project/cellpop/) and a JavaScript package on npm (https://www.npmjs.com/package/cellpop). The Python package provides a Jupyter widget implemented with anywidget [16].

Scellop is implemented in React, using *visx* (https://airbnb.io/visx/) to incorporate D3-based visualizations [17] for various scales and axis rendering. Undo and redo is supported through the *Zustand* state manager (https://zustand.docs.pmnd.rs/) with *zundo* middleware (http://github.com/charkour/zundo). All visualization panels can be resized, and the heatmap allows for zooming in on rows and columns. A configuration panel allows users to select theme and color schemes, set normalizations and transformations, determine side panels (bars, stacked bars, or violin plots), transpose the heatmap, and set different zoom levels. Users can select rows to display as bar charts embedded into the heatmap. Users can sort by counts, alphabetical, or sample and cell type metadata, such as donor age or cell ontology hierarchy. Data can be filtered based on related metadata values. Colors can also be configured individually for samples and cell types for bar charts and side panels. The resulting visualization can be exported as a high-resolution png. All interactions are also available from the context menu. By transposing the view, removing the heatmap, and using stacked bars in the side panel, the traditional stacked bar charts cell population plot can be created. This configuration can be accessed instantly as a preset from the settings. Data loading from (zarr-indexed) AnnData is supported, making this an scverse-compatible visualization tool. The Python package includes additional data loading functionality, supporting Pandas DataFrames and various ways of supplying metadata.

A demo with kidney RNAseq datasets is available at https://scellop.netlify.app/. Scellop is integrated within the HuBMAP Data Portal [18] as a default visualization to show organ-level overviews of cell type populations (e.g., https://portal.hubmapconsortium.org/organs/kidney#cell-population-plot) and as a template Python analysis in the integrated JupyterLab analysis environment for HuBMAP members (Workspaces).

# Application

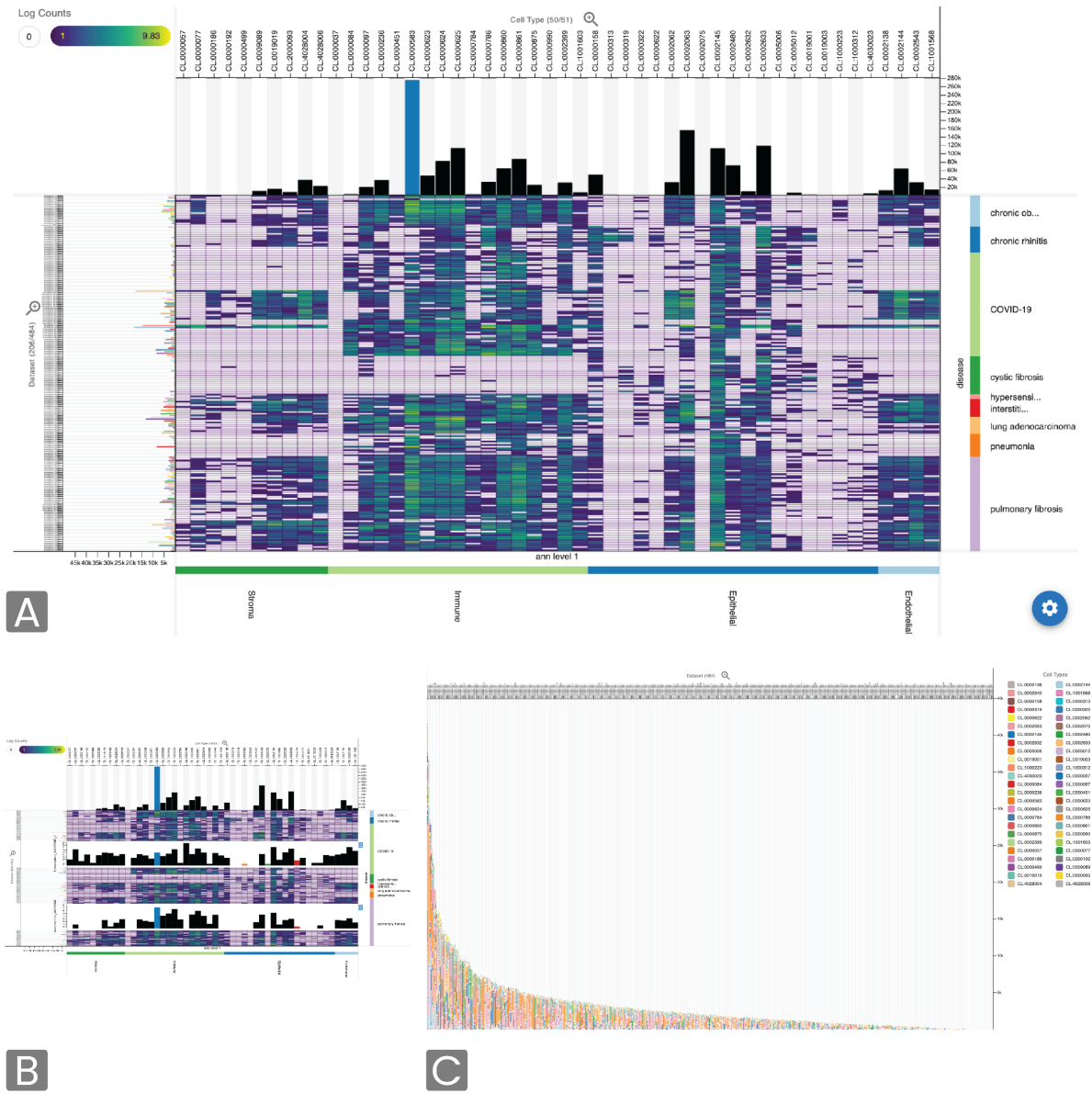

**Figure 1.** Scellop showing populations in Human Lung Cell Atlas datasets. A) Scellop view sorted by diseases and cell type annotation. B) Two datasets are converted to bar charts to compare. C) Stacked bar chart view.

Figure 1 shows the scellop viewer with data from the Human Lung Cell Atlas [19] (available from https://cellxgene.cziscience.com/collections/6f6d381a-7701-4781-935c-db10d30de293),
which constitutes of 484 datasets with 51 cell types (including 'unknown'). Scellop's default view shows a heatmap of cell counts with bar charts of total cell and sample counts on the side (Figure 1A). Sorting and filtering allow for grouping of cell types and datasets. Additional settings allow for log normalization and setting colors of bars, aiding communication of interesting results. It shows that donors with cystic fibrosis have a different cell type population, especially for immune cells. Certain datasets of donors with covid have different populations, so we can expand these datasets to compare them to different diseases (Figure 1B). The traditional stacked bar chart view can also be shown (Figure 1C). The amount of datasets complicates comparisons, especially those separated by distance [7]. Absent cell types and small fractions are hard to directly see, compared to scellop's main view. Individual datasets cannot be examined in a detailed view. Although datasets can be grouped in the stacked bar chart view, cell types hierarchies cannot be indicated. Because the scellop viewer can sort by subsequent hierarchy orders, we can show cell type relations. Thus, scellop is better suited for the identified user tasks than the commonly used stacked bar charts. Overall, the redesigned visualization approach proposed here allows for more granular exploration of cell populations. To facilitate inclusion in presentations and manuscripts, scellop views can also be exported as high-resolution images.

## Discussion

Scellop allows users to better explore cell populations with its interactive viewer, with easier pattern detection with its heatmap overview, and increased accuracy in comparing populations with its expandable bar charts. It better supports all of the user tasks identified via the design study, and can be integrated in Python and web environments for easy usage. Scellop can also be used in other domains where stacked bar chart visualizations are prevalent, such as in metagenomics [20,21]. A potential extension of scellop can include a network graph for hierarchical cell types, such as Collapsible Tree [22]. Sorting and filtering by hierarchies is supported by scellop, showing their relation can aid the exploration process. Hierarchical features would also support datasets with different cell type granularities better. Additionally, while scellop supports the widely used AnnData data format, data loading options can be expanded to alternative file formats.

## Acknowledgements

This work was supported by the National Institute of Health (OT2 OD033758 to N.G.). The authors wish to thank Trevor Manz for their support with using anywidget, and Yan Ma for supporting the progress of this project.

# Supplementary Materials

## Unique Cell Type Calculation

For calculating the average number of cell types, we used 162 datasets from the Human BioMolecular Atlas Program (HuBMAP). We retrieved the cell types with scellop's data loaders, and calculated the number of unique cell types per dataset. We used the following datasets, identified by their HuBMAP ID:

| | | | | | |
|---|---|---|---|---|---|
| HBM964.BVWP.737 | HBM675.VBDH.688 | HBM846.KVCF.674 | HBM266.FTJN.632 | HBM546.RNHX.756 | HBM528.KNCB.488 |
| HBM729.TVMN.534 | HBM734.LFLC.264 | HBM232.MBNR.586 | HBM965.PSNC.855 | HBM482.MCCP.264 | HBM796.PCWD.863 |
| HBM752.KZCK.589 | HBM373.HCFG.722 | HBM467.RQDN.922 | HBM626.PHCW.834 | HBM528.DMSV.294 | HBM324.XBMF.465 |
| HBM628.QKGB.497 | HBM462.XQCR.933 | HBM456.CGDP.395 | HBM976.MRWH.263 | HBM979.VMDC.365 | HBM647.QDBG.936 |
| HBM938.GBST.823 | HBM936.MHTZ.834 | HBM444.PWKX.639 | HBM674.FLVW.576 | HBM299.VDWT.444 | HBM456.XDCK.572 |
| HBM892.JLFV.844 | HBM437.LCSH.956 | HBM775.CMGG.464 | HBM937.TWRN.355 | HBM334.DWWF.436 | HBM736.MNMD.453 |
| HBM532.KKRC.477 | HBM539.GJNB.784 | HBM759.CHJW.244 | HBM247.HLXR.494 | HBM969.PBMH.689 | HBM699.XBTD.684 |
| HBM735.FSBZ.626 | HBM679.RLJH.432 | HBM823.CNRW.484 | HBM597.PBJW.593 | HBM253.ZBGF.863 | HBM563.FFQJ.764 |
| HBM222.VQSW.335 | HBM847.MDSJ.826 | HBM547.SJSK.268 | HBM883.PHQS.523 | HBM629.GSHG.922 | HBM264.MJCH.639 |
| HBM634.JHVB.286 | HBM363.NTWP.766 | HBM439.BQLR.867 | HBM928.THDD.545 | HBM874.JPGB.398 | HBM368.JCBG.263 |
| HBM634.ZSHF.736 | HBM766.NZWP.682 | HBM648.DKQK.874 | HBM842.DDTX.473 | HBM476.ZLDJ.925 | HBM363.FVKP.935 |
| HBM595.LBXP.486 | HBM793.JDRF.289 | HBM633.LLDZ.679 | HBM949.PLLF.787 | HBM694.NXCN.368 | HBM938.WTSR.492 |
| HBM398.BLRW.228 | HBM976.LDTR.982 | HBM297.FDTX.382 | HBM459.KCST.593 | HBM779.FQMX.497 | HBM986.KFWG.239 |
| HBM292.GSZL.269 | HBM827.MJMM.447 | HBM522.VFGB.335 | HBM248.HPXX.584 | HBM445.HBRQ.488 | HBM782.HVML.355 |
| HBM845.SFMK.942 | HBM547.TFRR.794 | HBM787.XCSX.733 | HBM975.MVDK.648 | HBM892.CCDZ.345 | HBM834.SLQN.292 |
| HBM593.CLXN.573 | HBM269.GDLH.894 | HBM485.VKSZ.779 | HBM294.XZLM.256 | HBM933.JFFT.692 | HBM356.MDPN.792 |
| HBM757.KLKW.524 | HBM785.XFTT.663 | HBM425.GDJT.648 | HBM522.QXVG.468 | HBM522.FTFK.487 | HBM449.QGGL.994 |
| HBM745.FJML.722 | HBM688.RPFC.258 | HBM894.DMKD.525 | HBM573.JGLL.575 | HBM478.VWXX.362 | HBM573.JMXM.823 |
| HBM762.RPDR.282 | HBM727.DWPV.852 | HBM468.SSXX.967 | HBM887.DDJL.589 | HBM589.THRM.428 | HBM859.LTWK.468 |
| HBM874.PWHS.622 | HBM858.MFWR.937 | HBM982.DSNZ.722 | HBM972.LBGS.258 | HBM975.WQQQ.853 | HBM832.WTNH.257 |
| HBM967.LPHM.957 | HBM444.DXLZ.643 | HBM726.NFVH.245 | HBM362.DZVK.533 | HBM975.JGXC.665 | HBM883.DKXZ.574 |
| HBM864.CWHJ.963 | HBM929.VSJQ.633 | HBM456.GRCM.369 | HBM482.DKQF.747 | HBM582.CXXZ.438 | HBM339.BGVK.388 |
| HBM932.ZMRS.894 | HBM373.SCNK.776 | HBM625.BCND.537 | HBM343.XKRX.239 | HBM735.NMFW.852 | HBM293.QVMW.765 |
| HBM487.MCTL.254 | HBM873.PZTG.367 | HBM529.KHGN.262 | HBM265.FQWZ.384 | HBM675.RVGB.258 | HBM798.BBXD.333 |
| HBM269.XWMK.444 | HBM384.FLVW.984 | HBM778.JJDB.736 | HBM326.WHVS.274 | HBM948.GXMD.986 | HBM578.BDBP.672 |
| HBM673.GSSW.364 | HBM492.CQJD.323 | HBM473.RKXT.944 | HBM982.THGM.772 | HBM342.CMHT.948 | HBM599.GNJN.777 |
| HBM957.TXXZ.387 | HBM385.LQVK.975 | HBM927.DZCV.762 | HBM537.MVDQ.934 | HBM236.JPVT.769 | HBM852.VXGN.375 |

## User Study

We conducted a user study on cell population plots as part of the HuBMAP Data Portal. Fourteen participants participated in 30 minute semi-structured interviews focused on cell population visualizations. Participants were recruited for their domain expertise in single-cell and spatial biology, with 12 self-identifying as experimental biologists, 5 computational biologists, 5 educators, 1 clinician, and 12 data contributors to HuBMAP.

Users were first asked about their desired visualizations and interactions for exploring cell types prior to being shown any example visualization. They were then asked to examine the cell population plot available at https://hubmapconsortium.github.io/tissue-bar-graphs/ (with 'Kidney HuBMAP Portal (107 datasets)' as source, and default sorts and groupings), also shown in Figure S1.

Users highlighted the importance of comparison of datasets and cell types. Users wanted to find relations between cell type distributions and metadata, specifically within demographic groups and with different cell types. Users also wanted to see the total number of cells, the most frequent cell type, and find the cell types with highest and lowest cell counts between datasets.

The desired interactions raised by multiple users were: Normalization Option [N=10], Group by Cell Type Hierarchy [N=9], Overview with Detail [N=9], Ability to manipulate visualization (e.g. selecting substacks, filtering, etc.) [N=8], Additional Context [N=5], Group samples by desired feature [N=5], Filter by cell types [N=4], Filter by data types [N=4], Filter by donor metadata [N=4], Filter by anatomical structures [N=3], Filter by biomarkers [N=2]. Issues raised by multiple users were: Color Scheme [N=6].

Additional points raised were that there were too many cells, that it was hard to see the big picture, that cell types can vary by granularity. Additional interactions raised by individuals were cell type sorting by prevalence, seeing all samples at once, and downloading the graph as an image. One user specified that they normally looked at at least 30 datasets at one time.

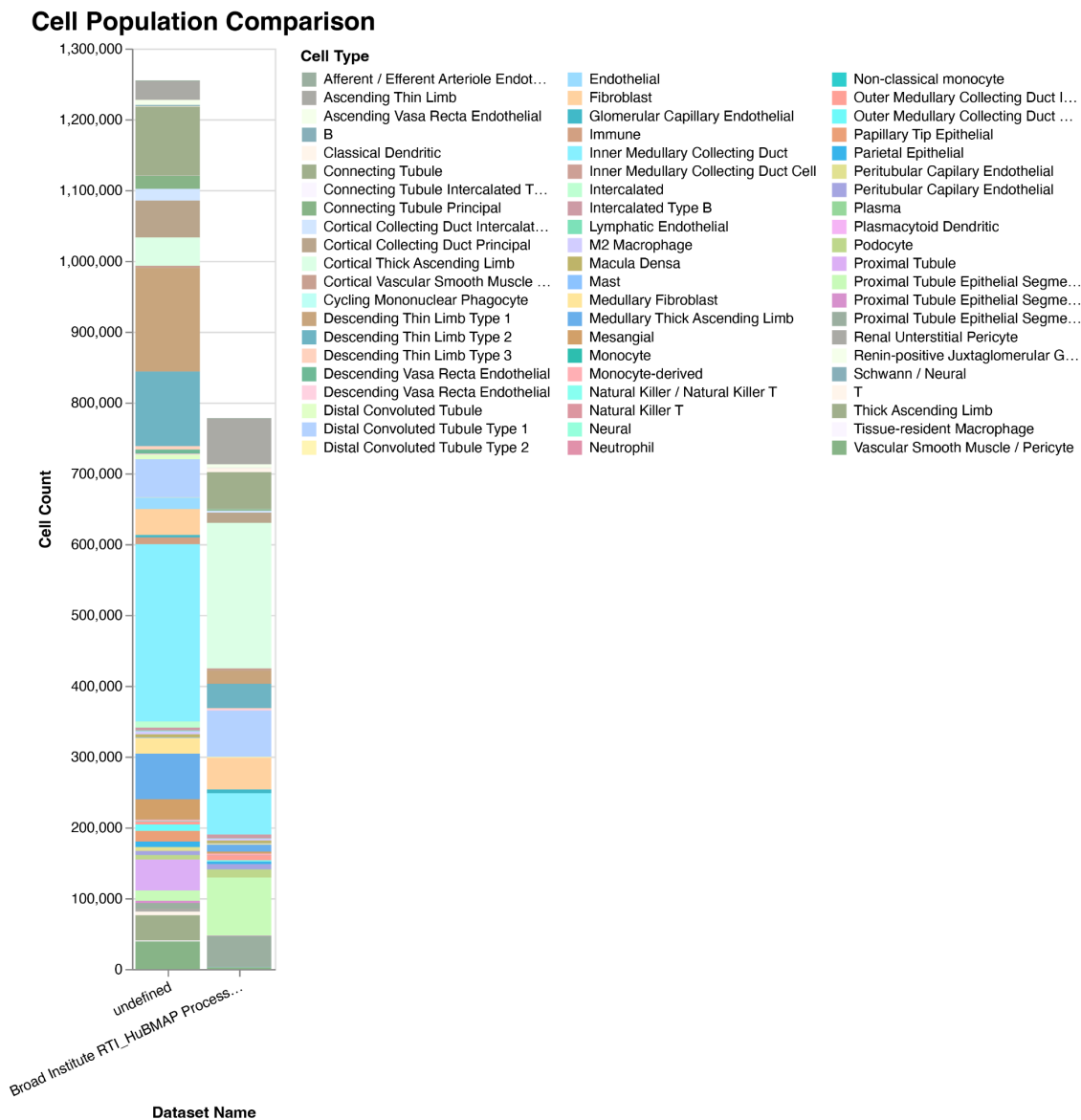

**Figure S1.** Cell Population Comparison used in HuBMAP Data Portal user study, with 63 cell types and two groups.